\documentclass[11pt]{article}

\usepackage{graphicx}
\usepackage{tikz-cd}
\usepackage{amssymb}

\newcommand{\Std}[1]{\mathbf{Std}(#1)}

\begin{document}
\begin{center}
\begin{Large}
\begin{tabular}{c}
Five Basic Concepts\\
of Axiomatic Rewriting Theory
\end{tabular}
\end{Large}

\medbreak
\medbreak

\begin{tabular}{c}
Paul-Andr\'e Melli\`es
\\
Institut de Recherche en Informatique Fondamentale (IRIF)
\\
CNRS, Universit\'e Paris Diderot
\end{tabular}
\end{center}

\vspace{.2em}

\begin{abstract}
In this invited talk, I will review five basic concepts of Axiomatic Rewriting Theory,
an axiomatic and diagrammatic theory of rewriting
started 25 years ago in a LICS paper with Georges Gonthier and Jean-Jacques L\'evy,
and developed along the subsequent years into a full-fledged 2-dimensional
theory of causality and residuation in rewriting.
I will give a contemporary view on the theory, informed by my later work on categorical semantics
and higher-dimensional algebra, and also indicate a number of current research directions in the field.
\end{abstract}

\vspace{1em}

A good way to understand Axiomatic Rewriting Theory 
is to think of it as a 2-dimensional refinement
of Abstract Rewriting Theory.
Recall that an abstract rewriting system 
is defined as a set~$V$ of vertices (= terms)
equipped with a binary relation ${\to}\subseteq V\times V$.
This abstract formulation is convenient to formulate various notions of termination and of confluence, 
and to compare them, typically:
\begin{center}
\begin{tabular}{ccc}
strong normalisation & vs. & weak normalisation
\\
confluence & vs. & local confluence
\end{tabular}
\end{center}
Unfortunately, the theory is not sufficiently informative
to capture more sophisticated structures and properties
of rewriting systems related to causality and residuation, like
\begin{center}
\begin{tabular}{c}
redexes and residuals
\\
finite developments
\\
standardisation
\\
head rewriting paths
\end{tabular}
\end{center}
These structures and properties are ubiquitous in rewriting theory.
They appear in conflict-free rewriting systems like the $\lambda$-calculus
as well as in rewriting systems with critical pairs, like action calculi and bigraphs
designed by Milner~\cite{milner} as universal calculus integrating the $\lambda$-calculus,
Petri nets and process calculi, or the $\lambda\sigma$-calculus introduced 
by Abadi, Cardelli, Curien and L\'evy \cite{abadi-cardelli-curien-levy} to express
in a single rewriting system the various evaluation strategies of an environment machine.

\medbreak

It thus makes sense to refine Abstract Rewriting Theory into a more sophisticated 
framework where the causal structures of computations could be studied for themselves,
in a generic way.
Intuitively, the causal structure of a rewriting path $f:M\twoheadrightarrow N$
is the cascade of elementary computations implemented by that path.
In order to extract these elementary computations from the rewriting path~$f$,
one needs to trace operations (= redexes) inside it.
This is achieved by permuting the order of execution of independent redexes
executed by~$f$.
An axiomatic rewriting system is thus defined as a graph
$G=(V,E,\partial_0,\partial_1)$ 
consisting of a set~$V$ of vertices (= the terms),
a set~$E$ of edges (= the redexes)
and a pair of source and target functions 
$\partial_0,\partial_1:E\to V$
equipped moreover with a family of permutation tiles,
satisfying a number of axiomatic properties.

\paragraph{1. Permutation tiles.}
The purpose of permutation tiles is to permute the order of execution of redexes.
In our axiomatic setting, a permutation tile $(f,g)$ is a pair of coinitial and cofinal
rewriting paths of the form: 
\begin{center}
\includegraphics{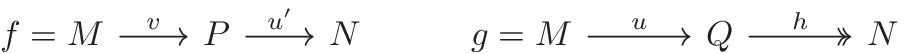}
\end{center}
where $u,v,u'$ are redexes and $h$ is a rewriting path.
The intuition is that $h$ computes the residuals of the redex~$v$ along the redex~$u$.
Two typical permutation tiles in the $\lambda$-calculus are the following one:
$$
\begin{array}{ccc}
\includegraphics[width=13em]{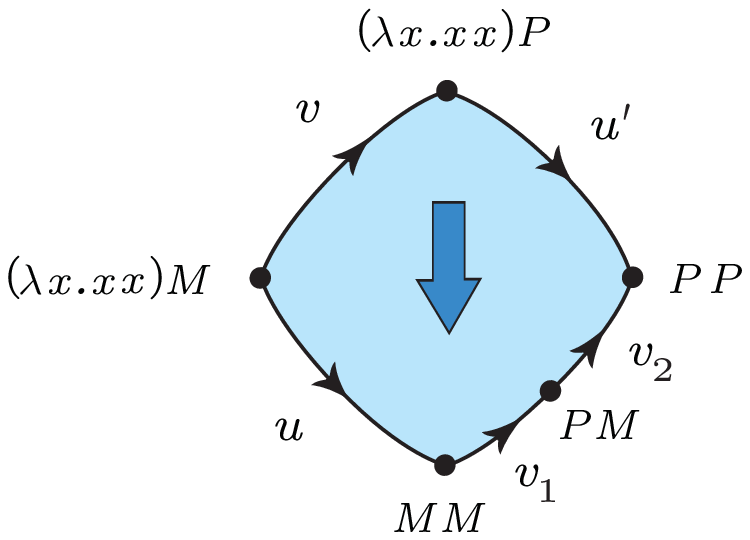}
& \quad &
\raisebox{1em}{\includegraphics[width=11em]{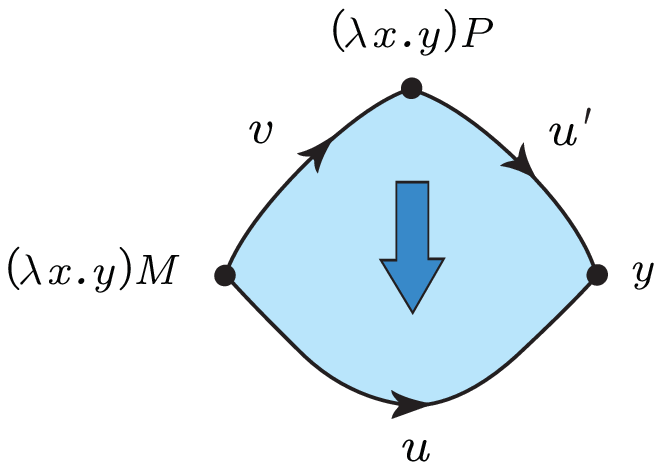}}
\end{array}
$$
where $h=v_1\cdot v_2$ on the left-hand side and $h=id$ on the right-hand side.

\paragraph{2. Standardisation cells.}
The permutation tiles are oriented, and generate a 2-dimensional rewriting system 
on the 1-dimensional rewriting paths.
In order to distinguish this rewriting system from the original rewriting system,
we call it the standardisation rewriting system.
A standardisation path~$\theta$ between 1-dimensional rewriting paths
$f,g:M\twoheadrightarrow N$ is then written as
$$
\theta \quad : \quad f \Rightarrow g \quad : \quad M \twoheadrightarrow N 
$$%
The axioms of Axiomatic Rewriting Theory are designed to ensure
that this 2-dimensional rewriting system is weakly normalising and confluent.
In order to establish weak normalisation, one needs to clarify an important point:
when should one consider that two standardisation paths
$$
\theta,\theta' \quad : \quad f \Rightarrow g \quad : \quad M \twoheadrightarrow N 
$$
are equal? The question looks a bit esoteric, but it is in fact fundamental!
By way of illustration, consider the following permutation tile in the $\lambda$-calculus:
\begin{equation}\label{equation/reversible-tile}
\raisebox{-4.3em}{\includegraphics[width=10em]{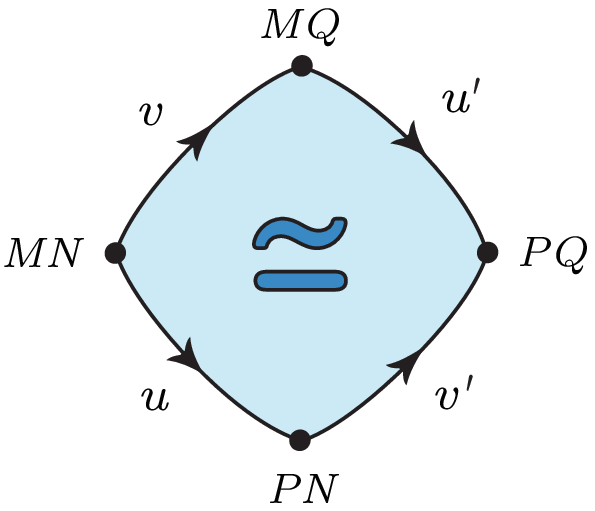}}
\end{equation}
where the two $\beta$-redexes $u$ and $v$ should be considered as syntactically disjoint
because $u$ is a $\beta$-redex of the subterm~$M$ and $v$ is a $\beta$-redex of the disjoint subterm~$N$.
If one does not want to give a left-to-right precedence to the $\beta$-redex~$u$
over the $\beta$-redex~$v$, one should equip the axiomatic rewriting system with two permutation tiles
$$
\theta_1:v\cdot u'\Rightarrow u\cdot v'
\quad\quad\quad\quad
\theta_2:u\cdot v'\Rightarrow v\cdot u'.
$$
The task of the permutation tile~$\theta_1$ is to permute~$u$ before~$v$,
while the task of the permutation tile~$\theta_2$ is to permute~$v$ before~$u$.
It thus makes sense to require that their composite are equal to the identity
in the standardisation rewriting system:
$$
\theta_1;\theta_2 = id : v\cdot u'\Rightarrow v\cdot u'
\quad\quad\quad\quad
\theta_2;\theta_1 = id: u\cdot v'\Rightarrow u\cdot v'
$$
Of course, this enforces that $\theta_1$ and $\theta_2$ are inverse.
One declares in that case that the permutation tile~(\ref{equation/reversible-tile}) is reversible.
A standardisation path~$\theta:f\Rightarrow g$ consisting only of such reversible permutation tiles 
is called reversible, and one writes $\theta:f\simeq g$ in that case.
A simple and elegant way to describe the equational theory on standardisation paths
is to equip every permutation tile $(f,g)$ with an ancestor function $\varphi:[n]\to[2]$
where $[k]=\{1,\dots,k\}$ and $n$ is the length of the path $g=u\cdot h$.
The purpose of the function~$\varphi$ is to map the index of redex 
in $g=u\cdot h$ to the index of its ancestor $f=v\cdot u'$,
in the following way:
\begin{center}
\includegraphics[width=11em]{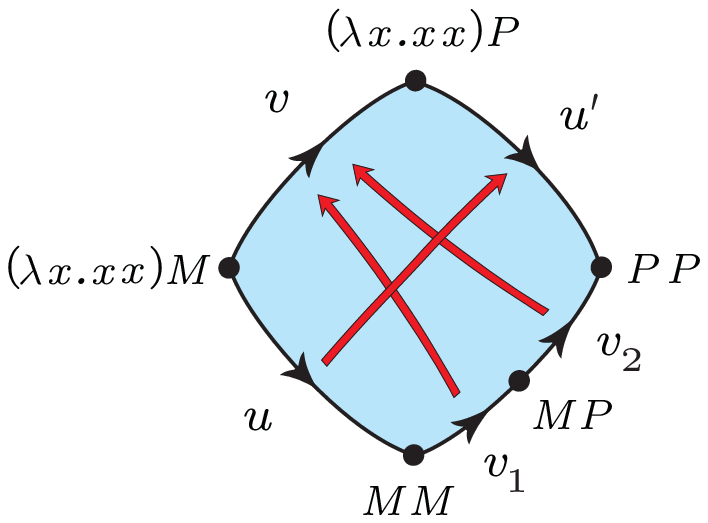}
\hspace{.5em}
\raisebox{1em}{\includegraphics[width=10em]{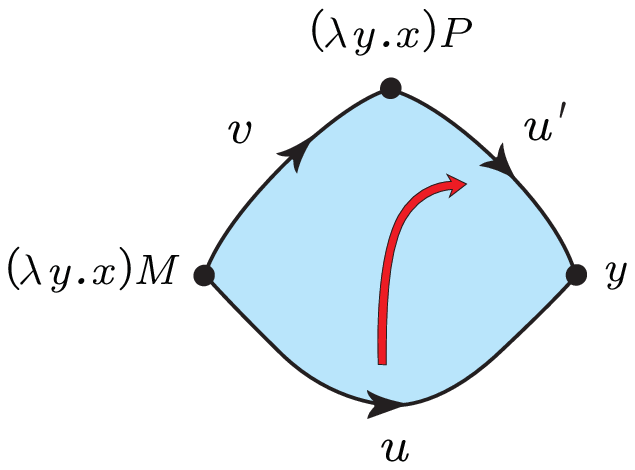}}
\hspace{.5em}
\raisebox{.3em}{\includegraphics[width=9em]{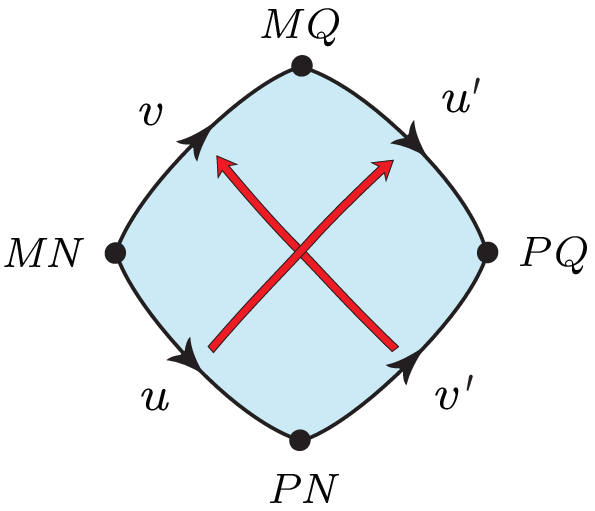}}
\end{center}
By way of illustration, the permutation tiles equipped with their ancestor functions 
may be composed in the following way in the $\lambda$-calculus:
\begin{center}
\includegraphics[width=20em]{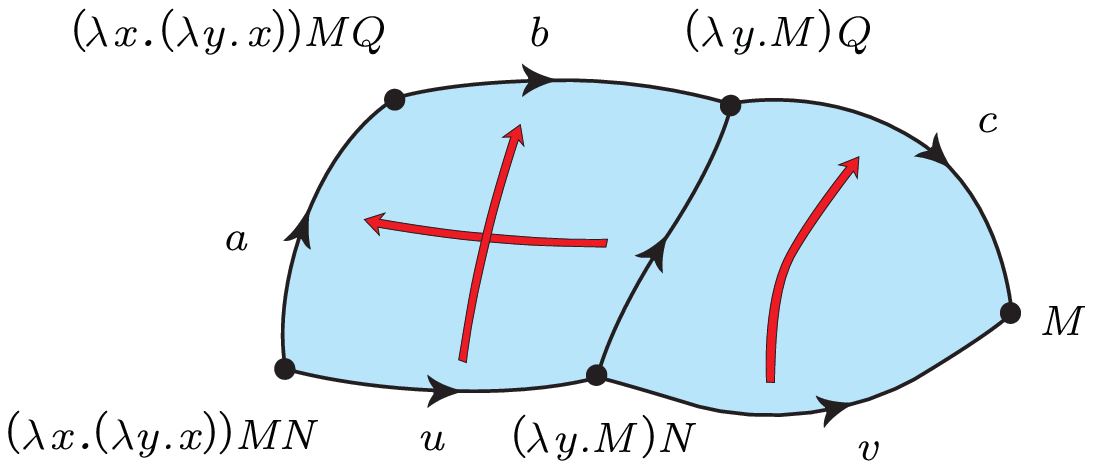}
\end{center}
This leads us to identify two standardisation paths $\theta,\theta':f\Rightarrow g$
when they produce the same ancestor function.
A standardisation cell is then defined as an equivalence class of standardisation paths
$\theta,\theta':f\Rightarrow g$ modulo this equivalence relation.
Note in particular that the equivalence relation identifies the standardisation path $\theta_1;\theta_2$ 
with the identity, and similarly for $\theta_2;\theta_1$.

\medbreak 

In this way, one defines for every axiomatic rewriting system~$G$
a 2-category $\Std{G}$ of whose objects are the vertices (= terms) of~$G$,
whose morphisms are the paths (= rewriting paths) of~$G$,
and whose 2-cells are the standardisation cells.
One declares that two rewriting paths~$f,g:M\twoheadrightarrow N$ are equivalent
modulo redex permutation (noted $f\sim g$) when $f$ and $g$ are in the same
connected component of the hom-category $\Std{G}(M,N)$ of rewriting paths from~$M$ to~$N$.
This means that one can construct a zig-zag of standardisation paths between~$f$ and~$g$.
%
We also like to say that the rewriting paths $f$ and $g$ are homotopy equivalent when $f\sim g$.

\paragraph{3. Standard rewriting paths.}
A rewriting path $f:M\twoheadrightarrow N$ is called standard when every standardisation cell
$\theta:f\Rightarrow g:M\twoheadrightarrow N$ is reversible.
The standardisation theorem states that 

\medbreak

\noindent
\emph{Standardisation Theorem.} For every rewriting path $f:M\twoheadrightarrow N$
there exists a standardisation cell $\theta:f\Rightarrow g$ to a standard rewriting path $g:M\twoheadrightarrow N$.
Moreover, this standard rewriting path is unique in the sense 
that for every standardisation cell $\theta':f\Rightarrow g'$ 
to a standard rewriting path~$g':M\twoheadrightarrow N$, there exists a reversible
standardisation path $\theta'':g'\simeq g$ such that $\theta=\theta';\theta''$.

\medbreak

\noindent
The theorem is established in any axiomatic rewriting system~$G$
using the elementary axioms on the permutation tiles provided by the theory.
As a matter of fact, the property is even stronger: it states that there exists 
a unique standardisation cell~$\theta$ from $f$ to the standard rewriting path~$g$.
This means that every standard path~$g:M\twoheadrightarrow N$ is a terminal object 
in its connected component of rewriting paths $f:M\twoheadrightarrow N$.
See \cite{mellies:phd,mellies:art1,mellies:hdr} for details.

\paragraph{4. External rewriting paths.}
An external rewriting path $e:M\twoheadrightarrow N$
is defined as a rewriting path
such that for every standard rewriting path $f:N\twoheadrightarrow P$, 
the composite rewriting path $e\cdot f:M\twoheadrightarrow P$ is standard.
Note in particular that every external rewriting path is standard.
Accordingly, a rewriting path $m:M\twoheadrightarrow N$ is called internal 
when for every standardisation cell $\theta:m\Rightarrow e\cdot f$
where the rewriting path~$e$ is external, the rewriting path~$e$
is in fact the identity on $M$.
One establishes the following property in every axiomatic rewriting system,
see \cite{mellies:art3} for details:

\medbreak

\noindent
\emph{Factorisation Theorem}:
For every rewriting path $f:M\twoheadrightarrow N$,
there exists a unique external rewriting path $e:M\twoheadrightarrow P$
and a unique internal rewriting path $m:P\twoheadrightarrow N$
up to permutation equivalence such that $f\sim e\cdot m$.
This factorization is moreover functorial.

\paragraph{5. Head-rewriting paths.}
The factorization theorem is supported by the intuition that only the external part
$e:M\twoheadrightarrow P$ of a rewriting path $f:M\twoheadrightarrow N$
performs relevant computations, while the internal part $m:P\twoheadrightarrow N$
produces essentially useless extra computations.
The factorization property plays a fundamental role in the theory.
In particular, it enables us to establish a stability theorem which shows 
the existence of head-rewriting paths in every axiomatic rewriting system,
even the rewriting system is non-deterministic and has critical pairs.
The stability theorem states that under very general and natural assumptions 
on a set $\mathcal{H}$ of head-values, see~\cite{mellies:art4}, the following property holds:

\medbreak

\noindent
\emph{Stability Theorem}:
For every term~$M$ of the axiomatic rewriting system,
there exists a cone of external paths (called head-rewriting paths)
$$
e_i \quad : \quad M \twoheadrightarrow V_i
\quad\quad\quad\quad \mbox{with} \,\, V_i\in\mathcal{H}
$$
indexed by $i\in I$, which satisfies the following universality property:
for every rewriting path $f:M\twoheadrightarrow W$ 
reaching a head-value $W\in\mathcal{H}$, there exists a unique index $i\in I$ 
such that the rewriting path $f$ factors as
$$
f  \sim  e_i\cdot h \quad : \quad  M\twoheadrightarrow W
$$
for a given rewriting path $h:V_i\twoheadrightarrow W$.
The rewriting path $h:V_i\twoheadrightarrow W$ is moreover unique 
modulo permutation equivalence.
In the case of axiomatic rewriting systems without critical pairs,
the theorem establishes the existence of a head-rewriting path $e:M\twoheadrightarrow V$
for every term $M$ which can be rewritten to a head-value $W\in\mathcal{H}$.
The stability theorem is particularly useful in rewriting systems with critical pairs.
By way of illustration, it enables one to describe the head-rewriting paths
$e_i:M\twoheadrightarrow V_i$ which transport a $\lambda$-term $M$ 
to its head-normal forms in the $\lambda\sigma$-calculus, see \cite{mellies:art2} for details.

\begin{footnotesize}

\end{footnotesize}

\end{document}